\PassOptionsToPackage{unicode}{hyperref}
\PassOptionsToPackage{hyphens}{url}
\PassOptionsToPackage{dvipsnames,svgnames,x11names}{xcolor}
\documentclass[
  12pt]{article}

\usepackage{amsmath,amssymb}
\usepackage{iftex}
\ifPDFTeX
  \usepackage[T1]{fontenc}
  \usepackage[utf8]{inputenc}
  \usepackage{textcomp} 
\else 
  \usepackage{unicode-math}
  \defaultfontfeatures{Scale=MatchLowercase}
  \defaultfontfeatures[\rmfamily]{Ligatures=TeX,Scale=1}
\fi
\usepackage{lmodern}
\ifPDFTeX\else  
\fi
\IfFileExists{upquote.sty}{\usepackage{upquote}}{}
\IfFileExists{microtype.sty}{
  \usepackage[]{microtype}
  \UseMicrotypeSet[protrusion]{basicmath} 
}{}
\makeatletter
\@ifundefined{KOMAClassName}{
  \IfFileExists{parskip.sty}{%
    \usepackage{parskip}
  }{
    \setlength{\parindent}{0pt}
    \setlength{\parskip}{6pt plus 2pt minus 1pt}}
}{
  \KOMAoptions{parskip=half}}
\makeatother
\usepackage{xcolor}
\makeatletter
\ifx\paragraph\undefined\else
  \let\oldparagraph\paragraph
  \renewcommand{\paragraph}{
    \@ifstar
      \xxxParagraphStar
      \xxxParagraphNoStar
  }
  \newcommand{\xxxParagraphStar}[1]{\oldparagraph*{#1}\mbox{}}
  \newcommand{\xxxParagraphNoStar}[1]{\oldparagraph{#1}\mbox{}}
\fi
\ifx\subparagraph\undefined\else
  \let\oldsubparagraph\subparagraph
  \renewcommand{\subparagraph}{
    \@ifstar
      \xxxSubParagraphStar
      \xxxSubParagraphNoStar
  }
  \newcommand{\xxxSubParagraphStar}[1]{\oldsubparagraph*{#1}\mbox{}}
  \newcommand{\xxxSubParagraphNoStar}[1]{\oldsubparagraph{#1}\mbox{}}
\fi
\makeatother

\usepackage{longtable,booktabs,array}
\usepackage{calc} 
\usepackage{etoolbox}
\makeatletter
\patchcmd\longtable{\par}{\if@noskipsec\mbox{}\fi\par}{}{}
\makeatother
\IfFileExists{footnotehyper.sty}{\usepackage{footnotehyper}}{\usepackage{footnote}}
\makesavenoteenv{longtable}
\usepackage{graphicx}
\makeatletter
\def\maxwidth{\ifdim\Gin@nat@width>\linewidth\linewidth\else\Gin@nat@width\fi}
\def\maxheight{\ifdim\Gin@nat@height>\textheight\textheight\else\Gin@nat@height\fi}
\makeatother
\setkeys{Gin}{width=\maxwidth,height=\maxheight,keepaspectratio}
\makeatletter
\def\fps@figure{htbp}
\makeatother

\makeatletter
\@ifpackageloaded{caption}{}{\usepackage{caption}}
\AtBeginDocument{%
\ifdefined\contentsname
  \renewcommand*\contentsname{Table of contents}
\else
  \newcommand\contentsname{Table of contents}
\fi
\ifdefined\listfigurename
  \renewcommand*\listfigurename{List of Figures}
\else
  \newcommand\listfigurename{List of Figures}
\fi
\ifdefined\listtablename
  \renewcommand*\listtablename{List of Tables}
\else
  \newcommand\listtablename{List of Tables}
\fi
\ifdefined\figurename
  \renewcommand*\figurename{Figure}
\else
  \newcommand\figurename{Figure}
\fi
\ifdefined\tablename
  \renewcommand*\tablename{Table}
\else
  \newcommand\tablename{Table}
\fi
}
\@ifpackageloaded{float}{}{\usepackage{float}}
\floatstyle{ruled}
\@ifundefined{c@chapter}{\newfloat{codelisting}{h}{lop}}{\newfloat{codelisting}{h}{lop}[chapter]}
\floatname{codelisting}{Listing}

\makeatother
\makeatletter
\@ifpackageloaded{caption}{}{\usepackage{caption}}
\@ifpackageloaded{subcaption}{}{\usepackage{subcaption}}
\makeatother

\ifLuaTeX
  \usepackage{selnolig}  
\fi
\usepackage[]{natbib}
\usepackage{bookmark}

\IfFileExists{xurl.sty}{\usepackage{xurl}}{} 
\hypersetup{
  pdftitle={Title},
  pdfauthor={Author 1; Author 2},
  pdfkeywords={3 to 6 keywords, that do not appear in the title},
  colorlinks=true,
  linkcolor={blue},
  filecolor={Maroon},
  citecolor={Blue},
  urlcolor={Blue},
  pdfcreator={LaTeX via pandoc}}

\newcommand{\anon}{1}

\begin{document}

\def\spacingset#1{\renewcommand{\baselinestretch}%
{#1}\small\normalsize} \spacingset{1}


\if1\anon
{
  \title{\bf Bias Correction for Scalar-on-Density Regression Models}
  \author{Fenglin Xie\thanks{CONTACT: fx2212@cumc.columbia.edu}\hspace{.2cm}\\
    Department of Biostatistics, Columbia University\\
    and \\
    Todd Ogden \\
    Department of Biostatistics, Columbia University}
  \maketitle
} \fi

\if0\anon
{
  \bigskip
  \bigskip
  \bigskip
  \begin{center}
    {\LARGE\bf Bias Correction for Scalar-on-Density Regression Models}
\end{center}
  \medskip
} \fi

\bigskip
\begin{abstract}
In scalar-on-density regression, the covariate is a density that is estimated from a finite number of measurements gathered for each observational unit. When this number of measurements is relatively small, the estimated coefficient function suffers from attenuation bias. This paper studies how the bias depends on the number of measurements per unit and proposes a bias‑correction method based on simulation extrapolation (SIMEX). We establish that the bias decreases monotonically as the number of measurements per unit increases. The proposed SIMEX procedure applies bootstrap resampling to simulate smaller measurement counts and then extrapolates to infinitely many measurements, thereby correcting finite‑measurement bias. A comprehensive simulation study, conducted over a range of sample sizes and noise levels, shows that the mean integrated squared error of the coefficient function decreases with more measurements per unit and that the SIMEX‑extrapolated estimates achieve lower bias than the naive estimates based on the full set of measurements. The practical utility of the method is further illustrated through an application to the National Health and Nutrition Examination Survey, for which we relate 24‑hour physical activity profiles to all‑cause mortality. This example supports the validity of the method and demonstrates its ability to detect and correct for finite‑measurement bias.
\end{abstract}

\noindent%
\vfill

\newpage
\spacingset{1.8} 

\section{Introduction}
\label{s:intro}

Scalar-on-function regression (SoFR) provides a flexible framework for modeling the relationship between a scalar outcome and a functional predictor. Early work on SoFR includes basis expansion techniques \citep{cardot1999functional}, penalized spline methods \citep{marx1999generalized}, and functional principal component regression \citep{reiss2010functional}.  A comprehensive review of SoFR methods is given by \cite{reiss2017methods}. These methods have been successfully applied in diverse fields, including neuroscience, epidemiology, and personalized medicine.

In many contemporary settings, the functional predictor is not a directly observed smooth curve but rather a density estimated from repeated measurements per observational unit. For instance, in clinical research, the distribution of physical activity intensities measured by wearable accelerometers can predict disability scores in multiple sclerosis patients \citep{niyogi2026scalar}. In agricultural economics, the full distribution of daily temperatures over a growing season can be used to predict rice yield \citep{trinh2026scalar}.  A natural extension of SoFR, this is known as scalar-on-density regression. A straightforward approach is to estimate each subject’s density nonparametrically and then treat the estimated density as a functional covariate in a standard SoFR model. However, the space of density functions is not a linear space, and so direct application of linear functional regression may be inappropriate. A common remedy is to transform the density to a Hilbert space, for example using the log‑quantile density transformation, so that standard linear models can be applied in the transformed domain \citep{petersen2016functional, tang2023differences}.

In classical regression with scalar or vector predictors, measurement error in the covariates causes attenuation bias, shrinking the estimated coefficients toward zero, and the magnitude of this bias decreases as the number of repeated measurements increases \citep{fuller2009measurement}. 

A similar attenuation phenomenon occurs in functional regression when the functional covariate is observed with error. Several recent works have addressed this problem. 
\cite{cai2015methods} introduced SIMEX-based approaches for bias correction in SoFR modeling when the functional covariates are observed with additive noise.
\cite{chen2024adjusting} proposed SIMEX-based procedures for functional quantile regression with error-prone covariates; and \cite{luan2023generalized} proposed SIMEX and regression calibration methods for generalized functional linear regression with mixture of error-prone functional and scalar covariates. Each of these studies assumes additive noise in the functional predictors.  

In scalar-on-density regression, the accuracy of the estimation of the density for each subject depends primarily on
the number of measurements made per unit. When this number is relatively small, there is more uncertainty in the estimation of the density, which can lead to attenuation bias when estimating the   coefficient function. As the number of measurements per unit increases, the bias diminishes.

The contribution of this paper includes adaptation of the SIMEX methodology to the scalar-on-density setting, in which the errors associated with the functional predictors is not additive but is instead a natural consequence of density estimation with a finite number of measurements per observational unit. Applying the general principle of SIMEX methods, we can correct the bias in estimation of the coefficient function by bootstrapping smaller numbers of measurements per unit and then extrapolating estimates to the limit represented by perfectly estimated densities.  

The remainder of the paper is organized as follows. Section 2 presents the proposed SIMEX-based bias correction method. Section 3 describes the simulation design and results. Section 4 illustrates the method using physical activity data from the National Health and Nutrition Examination Survey (NHANES). Section 5 concludes with a discussion.

\section{Methods}

\subsection{Scalar-on-Function Regression Model}

Let \(Y_i\) denote a scalar outcome for unit \(i\), and let \(X_i(t)\) be a smooth function of \(t\in\mathcal{T}\) representing the functional covariate. The classical scalar-on-function regression (SoFR) model is given by
\begin{equation}\label{SoFR_model}
Y_i = \alpha + \int_{\mathcal{T}} X_i(t)\gamma(t)\,dt + \varepsilon_i,\qquad i=1,\dots,N,
\end{equation}
where \(\alpha\) is a scalar-valued intercept, \(\gamma(\cdot)\) is the unknown coefficient function, and \(\varepsilon_i\) are independent zero-mean errors with finite variance. The goal is to estimate \(\gamma\) and \(\alpha\) from an observed sample of \(N\) units.

\subsection{Scalar-on-Density Regression and the Need for Transformation}

In many applications, the functional predictor is not directly observed as a smooth curve but rather as a density estimated from a collection of  measurements per unit. For example, in one approach to the analysis of functional connectivity data, the distribution of correlation coefficients between brain regions serves as a predictor. For each subject, the collection of all pairwise correlation coefficients between brain regions, thereby summarizing the global functional connectivity architecture, forms an empirical distribution which can be used to predict some scalar outcome\citep{tang2023differences}. 

Denote by $f_i$ the true density of the measurements for the $i$th observational unit. A natural approach would be to simply use the density function $f_i(t)$ stand in for the functional covariate $X_i(t)$ in equation~\ref{SoFR_model} 
and applying standard SoFR methods.   
However, the space of density functions is not a linear space: densities are nonnegative and integrate to one, and the set of all densities does not form a vector space under usual addition and scalar multiplication. 

A standard solution is to transform space of density functions to  a Hilbert space. One common transformation is the log‑quantile density (LQD) transformation \citep{petersen2016functional}, defined as \(g_i(q) = \log\bigl\{f_i(F_i^{-1}(q))\bigr\}\), where \(F_i\) is the cumulative distribution function corresponding to \(f_i\) and \(q\in[0,1]\). The LQD transformation maps densities into the space of square‑integrable functions on \([0,1]\). After applying this transformation, we can fit a linear model with \(g_i(q)\) as the functional covariate:
\begin{equation}\label{LQD_model}
Y_i = \alpha + \int_0^1 g_i(q)\,\beta(q)\,dq + \varepsilon_i,
\end{equation}
where \(\beta(\cdot)\) is the coefficient function associated with the LQD-transformed predictor. Model~(\ref{LQD_model}) is now a standard scalar-on-function regression of the model~(\ref{SoFR_model}).

\subsection{Attenuation Due to Uncertainty in Density Estimation}

In application, the density of each subject is not known and must be estimated from a finite number of  measurements. Denote by \(m_i\) the number of measurements available for each unit $i$. The estimated density \(\hat{f}_i\) is then used to compute the transformed covariate \(\hat{g}_i(q)\). Because \(\hat{f}_i\) is estimated from a finite sample, the resulting \(\hat{g}_i(q)\) is also measured with error. In any regression model, measurement error in the covariates induces attenuation bias: the estimated coefficient function is shrunk toward zero relative to the true function. It has been demonstrated that similar bias occurs in functional regression settings when the functional covariate is estimated with additive error \citep{cai2015methods,tekwe2022estimation, chen2024adjusting}. 

We illustrate the effect of bias induced by uncertainty in density estimation with a simple simulation study. For each observational unit, we generated a ``true'' density function that is a mixture of beta, normal, and uniform distributions (see Section~\ref{subsec:simulation_details} for a complete description). We then drew \(m_i\) = $m$ measurements from each unit's true density, estimated the density using smoothing splines (via the fda package), applied the LQD transformation, and finally fitted a scalar‑on‑function regression model using penalized splines (the \texttt{lf} method in the \texttt{pfr} function) to obtain an estimate of the coefficient function \(\beta(\cdot)\). For this illustration, we fixed  \(N = 100\) and 
$\textrm{Var}(\epsilon_i)=0.01^2$ 
in the model~(\ref{LQD_model}).

\begin{figure}[htbp]
\centering
\begin{minipage}{0.48\textwidth}
  \centering
  \includegraphics[width=\linewidth]{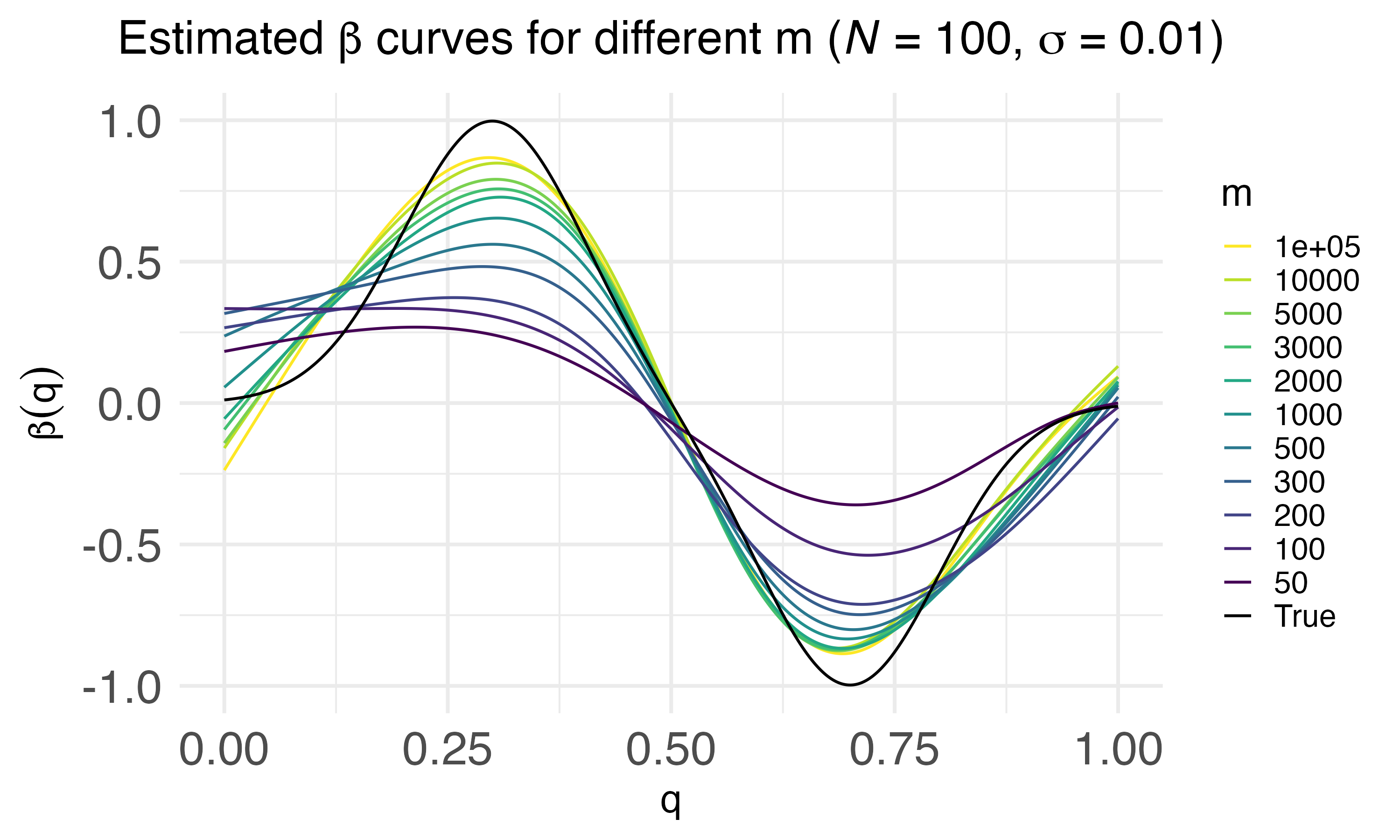}
  \vspace{0.2em}
  \par\small (a)
  \label{fig:beta_convergence}
\end{minipage}
\hfill
\begin{minipage}{0.48\textwidth}
  \centering
  \includegraphics[width=\linewidth]{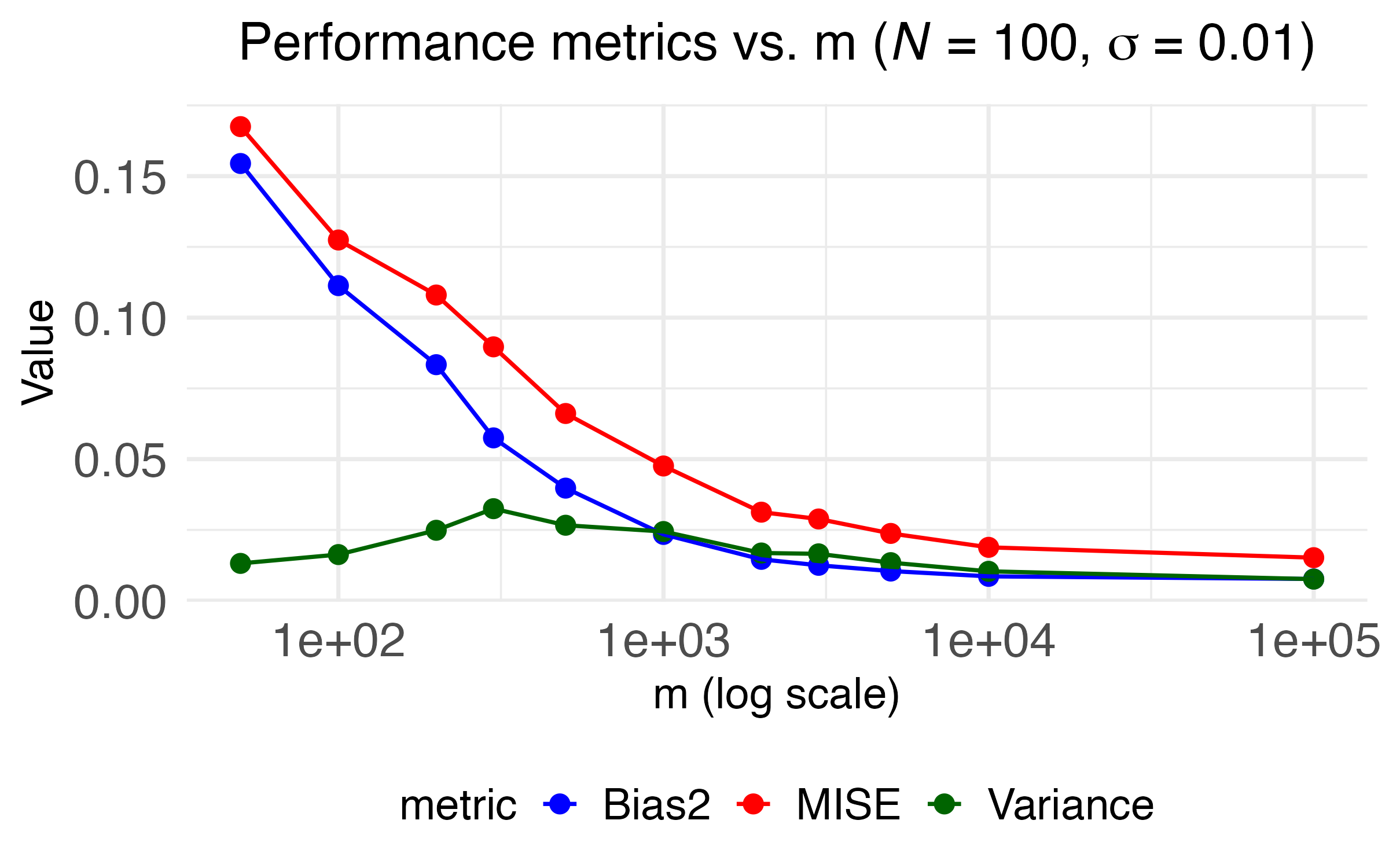}
  \vspace{0.2em}
  \par\small (b)
  \label{fig:metrics}
\end{minipage}
\caption{Behaviour of the off‑the‑shelf estimator in scalar‑on‑density regression with a finite number of measurements (simulation with \(N=100\) units and noise level \(\sigma=0.01\)). Panel (a) displays estimates of \(\beta(q)\) for increasing numbers of measurements \(m\) (coloured curves) getting closer to the true function (black). Panel (b) shows the mean integrated squared error, integrated squared bias and integrated variance of the estimator as functions of \(m\).}
\label{fig:convergence_and_metrics}
\end{figure}

Figure~\ref{fig:convergence_and_metrics} panel (a) displays means (over 100 independent datasets) of the estimates of \(\beta(\cdot)\) for various values of the per‑subject sample size \(m\) (ranging from \(50\) to \(100000\)). As \(m\) grows, the estimate nears the true function. Panel (b) shows the mean integrated squared error (MISE), integrated squared bias, and integrated variance of the off‑the‑shelf estimator, averaged over 100 independent datasets.  For a coefficient function estimate \(\hat\beta\), these quantities are defined as
\[
\mathrm{MISE}(\hat\beta) = \int_0^1 \bigl(\hat\beta(q) - \beta(q)\bigr)^2 dq,
\]
\[
\mathrm{Bias}^2 = \int_0^1 \bigl(E[\hat\beta(q)] - \beta(q)\bigr)^2 dq,
\]
\[
\mathrm{Variance} = \int_0^1 \mathrm{Var}\bigl(\hat\beta(q)\bigr) dq.
\]
The MISE and the integrated squared bias decrease monotonically with \(m\), with the variance remaining relatively constant, clearly illustrating the attenuation effect. This bias motivates the need for a correction method that can extrapolate to the ideal setting of infinitely many measurements per unit.

\subsection{Simulation Extrapolation (SIMEX) for SoFR}\label{subsec:sofr_simex}

The simulation extrapolation (SIMEX) method was originally developed for classical measurement error models with scalar covariates \citep{cook1994simulation}. The key idea is to add controlled amounts of additional measurement error to the data, study how the estimate changes as a function of the added error variance, and then extrapolate back to the zero‑error case.

In the context of SoFR with a functional covariate subject to additive measurement error, SIMEX proceeds as follows. First, an estimate of the measurement error variance $\sigma^2$  is obtained, for example by smoothing the observed curves or by exploiting replicated measurements. Then, for a grid of nonnegative values $\theta$ , where the added noise has variance $\theta \times \sigma^2$, pseudo-errors are repeatedly generated and added to the observed covariate functions, creating a sequence of noisier versions. For each simulated noisy dataset, the SoFR model is fitted, yielding an estimate $\hat\beta_{k,\theta}(t)$ for the $k$‑th simulation. The estimates are then averaged across the $K$ simulations to obtain $\bar{\hat\beta}_\theta(t) = K^{-1}\sum_{k=1}^K \hat\beta_{k,\theta}(t)$. This average varies smoothly with both $\theta$ and $t$, so the extrapolation method is appled
to a two‑dimensional surface. Finally, the relationship between $\bar{\hat\beta}_\theta(t)$ and $\theta$ is extrapolated to $\theta = -1$, which corresponds to the hypothetical situation of no measurement error. Various extrapolation techniques, such as linear or quadratic extrapolation, can be employed \citep{cai2015methods}.

\subsection{Proposed SIMEX Method for Scalar-on-Density Regression}\label{subsec:our_simex}

We now adapt SIMEX to the scalar‑on‑density setting, where the error is not additive but instead arises from estimating the density based on a finite number of measurements per unit. The key observation is that the effective error level can be controlled by taking bootstrap subsamples of smaller numbers of measurements and using these subsamples to re‑estimate the density. Specifically, given a dataset in which each unit has  \(m\) measurements. From this original set, we draw bootstrap samples of size \(m_0\) (with \(m_0 < m\)) for each unit, estimate the density based on this bootstrap sample, apply the LQD transformation, and refit the regression model. Repeating this process many times for each \(m_0\) gives an average coefficient function estimate \(\bar{\hat{\beta}}_{m_0}(\cdot)\). Because the estimation error decreases as \(m_0\) increases, the sequence \(\bar{\hat{\beta}}_{m_0}(\cdot)\) should get closer to the true \(\beta(\cdot)\) as \(m_0\to\infty\) (though actual convergence to $\beta$ will also require $N$ to go to infinity, among other conditions).

The proposed algorithm proceeds as follows given $m$ measurements for each unit: First, we choose a grid of subsamples of size \(m_{01} < m_{02} < \dots < m_{0k} < m\).  For each such \(m_0\), we perform \(B\) bootstrap iterations. In each iteration, we draw \(m_0\) measurements with replacement from each unit’s original measurements, estimate $f_i$, apply the LQD transformation to obtain \(\hat{g}_i\), and fit the scalar‑on‑function regression model with the transformed density as the functional covariate in model~(\ref{LQD_model}). The resulting estimated coefficient function is recorded as $\hat{\beta}^{(b)}_{m_0}(\cdot)$. Then we average the estimates to obtain the average estimated coefficient function $\bar{\hat{\beta}}_{m_0}(\cdot) = \frac{1}{B}\sum_{b=1}^B \hat{\beta}^{(b)}_{m_0}(\cdot)$.

In the SIMEX setting for scalar‑on‑function regression with a known measurement error variance, one adds controlled amounts of error controlled by a parameter \(\theta\ge 0\) and then extrapolates the resulting estimates to the hypothetical no‑error case \(\theta = -1\). When density functions are the source of the predictors, the measurement error is not additive but is inherent in the finite number of measurements used to estimate each unit’s density. The effective error level decreases as the bootstrap sample of size \(m_0\) increases, and vanishes only when \(m_0\to\infty\). To extrapolate to the infinite measurement limit, we treat the number of measurements \(m_0\) as a measure of the inverse error level and define a transformation \(\lambda = 1/\sqrt{m_0}\). For each fixed \(q\), the points \((\lambda, \bar{\hat{\beta}}_{m_0}(q))\) are used to extrapolate to \(\lambda = 0\), which corresponds to the limit of infinitely many measurements per unit as \(m_0\to\infty\). Extrapolation is performed using a nonlinear model, specifically a tensor product spline that smooths the two‑dimensional surface \(\bar{\hat{\beta}}(q,\lambda)\). The bias‑corrected SIMEX estimate \(\hat{\beta}_{\mathrm{SIMEX}}(\cdot)\) is the smoothed surface evaluated at \(\lambda = 0\) \citep{cai2015methods}.

\section{Simulation Study}

We conducted a simulation study to evaluate the performance of the proposed SIMEX procedure for scalar‑on‑density regression when the number of measurements per unit is finite. The goal was to assess how the integrated bias and MISE of the estimated coefficient function depend on the number of measurements per unit, and whether the SIMEX extrapolation can reduce the bias compared with the naive (uncorrected) estimator.

\subsection{Simulation Settings}\label{subsec:simulation_details}

The data were generated from model~(\ref{LQD_model}) with intercept \(\alpha = 1\) and iid errors \(\varepsilon_i \sim N(0,\sigma^2)\) where \(\sigma \in \{0.01,0.05\}\). The true coefficient function \(\beta\) was set to be
\begin{equation}\label{beta_def}
\beta(q) = 0.25\,\phi(q;0.3,0.1^2)-0.25\,\phi(q;0.7,0.1^2),
\end{equation}
with
\(\phi(\cdot;\mu,\sigma^2)\) the normal density. We subtracted the mean of the function in~(\ref{beta_def})  to ensure that
\(\int_0^1\beta(q)dq=0\). For each unit $i$, we generated a ``true'' density $f_i$ on $[0,1]$ as a mixture of a beta distribution, a normal distribution, and a uniform distribution, truncated to $[0,1]$ and renormalized. Specifically, we generated a random mixture weight $w_i \sim \mathrm{Uniform}(0.4,0.7)$, beta parameters $\alpha_i \sim \mathrm{Uniform}(1.8,2.5),\; \beta_i \sim \mathrm{Uniform}(4,6)$, and normal parameters $\mu_i \sim \mathrm{Uniform}(0.6,0.8),\; \sigma_i \sim \mathrm{Uniform}(0.08,0.15)$. The unnormalized mixture density is
\[
h_i(t) = 0.7\bigl[w_i\,\mathrm{Beta}(t;\alpha_i,\beta_i) + (1-w_i)\,\mathrm{Normal}(t;\mu_i,\sigma_i^2)\bigr] + 0.3\,\mathrm{Uniform}(t;0,1),
\]
where $\mathrm{Beta}$, $\mathrm{Normal}$, and $\mathrm{Uniform}$ denote the respective density functions. We then restrict $t$ to $[0,1]$ and renormalize:
\[
f_i(t) = \frac{h_i(t)}{\int_0^1 h_i(t)\,du}, \qquad t\in[0,1].
\]
The uniform component (weight $0.3$) was added to improve numerical stability near the boundaries.

A large pool of \(10\,000\) units was generated once. For each simulation replication, we randomly drew \(N \in \{50,100,200\}\) units from this pool, and for each selected unit we generated a fixed “full” sample of \(m = 1000\) independent measurements from its true density \(f_i\). For each combination of $N$ and $\sigma$ we generated $100$ independent datasets. For each dataset, we then applied the proposed SIMEX procedure using a grid of bootstrap samples of size $m_0 \in \{10,20,50,100,200,500,1000\}$.

For each value of \(m_0\) and each bootstrap replicate, we first drew a bootstrap sample of size \(m_0\) with replacement from the \(m = 1000\) fixed measurements of each unit. The density was then estimated using smoothing splines via the \texttt{fda::density.fd} function (with 19 $B$‑spline basis functions and a second‑order roughness penalty). The smoothing parameter was selected automatically by the function using generalized cross‑validation. The estimated density was then transformed to its log‑quantile density, which also truncates the tails to avoid numerical instability. The resulting functional covariate \(\hat g_i\) was used in the scalar‑on‑density regression model, which we fitted using the \texttt{pfr} function from the \texttt{refund} package with the \texttt{lf} term (linear functional predictor). We used \(k = 20\) cubic B‑spline basis functions with a second‑order difference penalty ($P$‑splines), and the smoothing parameter was chosen automatically by restricted maximum likelihood (REML). For a given \(m_0\), we repeated this process \(B = 100\) times, and averaged the \(B\) estimated coefficient functions to obtain average \(\bar{\hat{\beta}}_{m_0}(\cdot)\). This produced a sequence of average curves \(\bar{\hat{\beta}}_{m_0}(\cdot)\) for \(m_0 = 10,20,50,100,200,500,1000\).
For each parameter combination (\(N,\sigma\)), we generated 100 independent datasets. For each dataset, we applied three SIMEX extrapolation methods (linear, quadratic, and nonlinear) to the sequence \(\bar{\hat{\beta}}_{m_0}(\cdot)\) to obtain three SIMEX coefficient function estimates \citep{cai2015methods}. 

\subsection{Results}\label{sec:simulation_results}

\begin{figure}[htbp]
\centering
\includegraphics[width=0.9\textwidth]{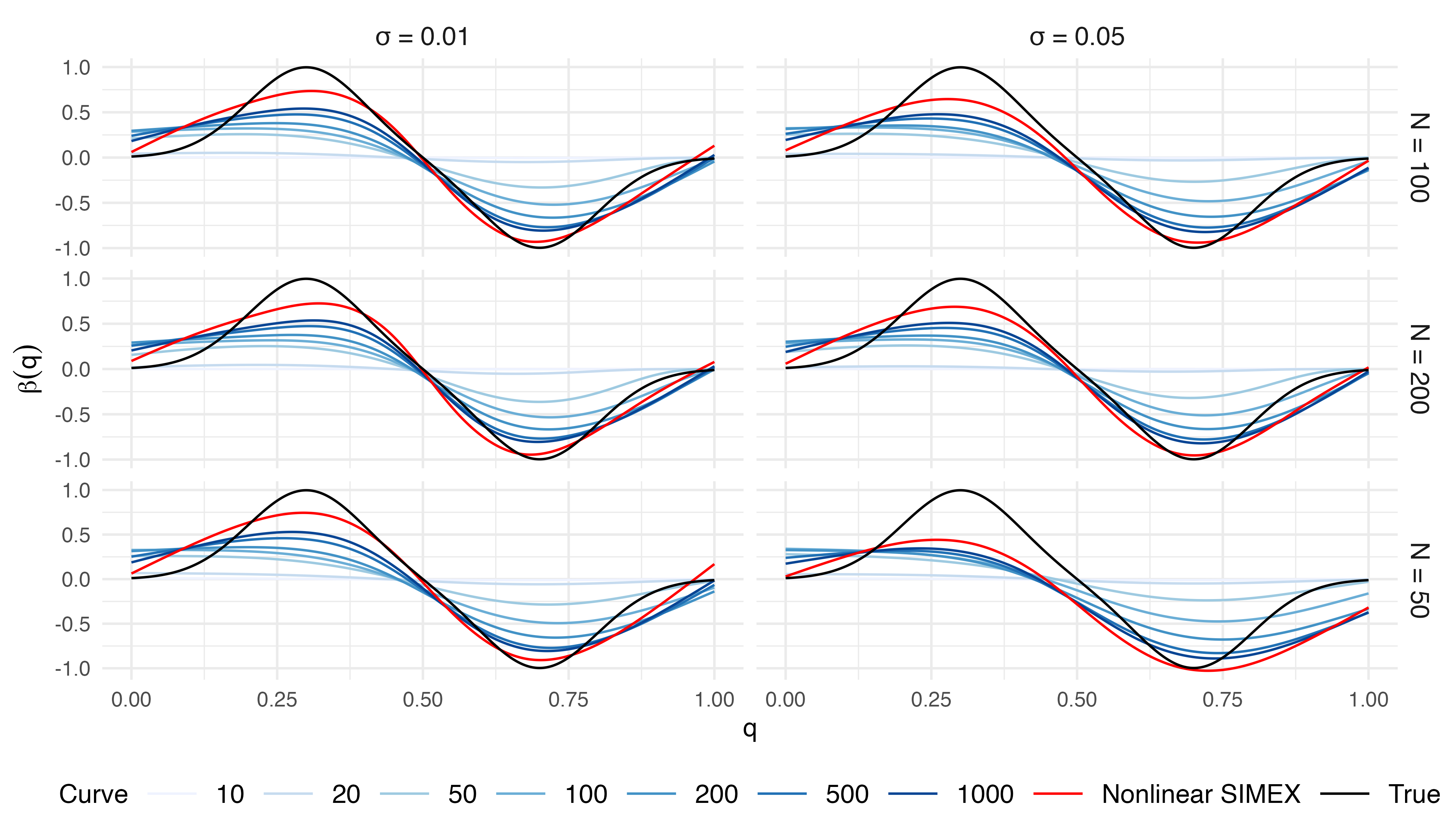}
\caption{Average estimated coefficient functions for all six parameter combinations. Blue curves (darker for larger \(m_0\)) represent the unadjusted estimates; red curves are the nonlinear SIMEX estimates; black curves are the true function.}
\label{fig:beta_curves}
\end{figure}

Figure~\ref{fig:beta_curves} displays the average estimated coefficient functions for the unadjusted estimator with increasing \(m_0\), the nonlinear SIMEX estimator, and the true beta function, for all six parameter combinations (\(N=50,100,200\) and \(\sigma=0.01,0.05\)). As \(m_0\) increases, the unadjusted curves get closer to the true function, and the SIMEX curve closely tracks the true shape, particularly around the peaks where the true coefficient function has its main features. For larger \(N\) and smaller \(\sigma\), the approach is faster and the SIMEX correction is more precise.

\begin{figure}[htbp]
\centering
\includegraphics[width=\textwidth]{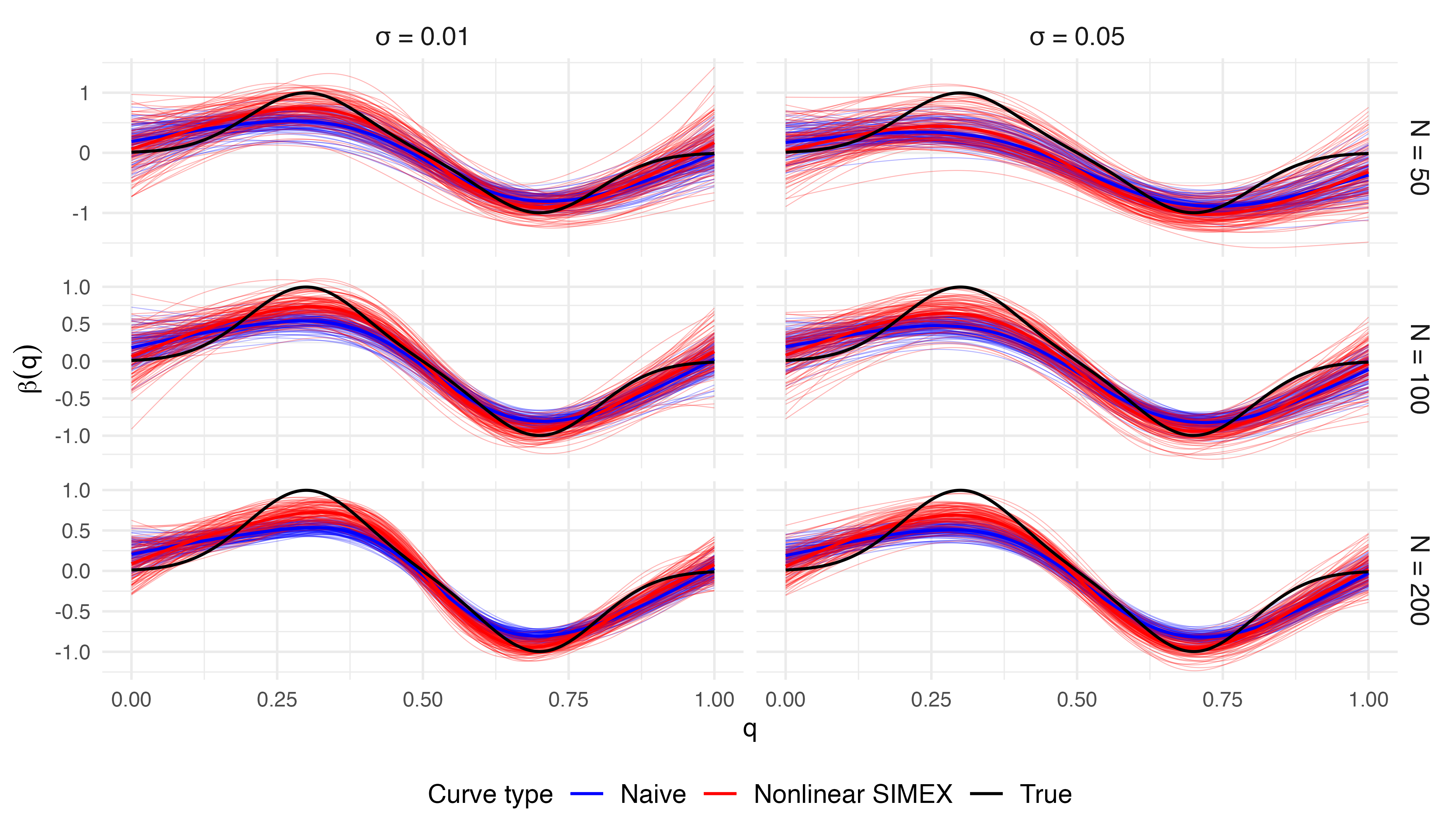}
\caption{Individual (light lines) and average (dark lines) naive (blue) and nonlinear SIMEX (red) coefficient function estimates across 100 replications, for each combination of sample size \(N\) and noise level \(\sigma\). The ``true'' function is shown in black.}
\label{fig:spaghetti_all}
\end{figure}

The spaghetti plots in Figure~\ref{fig:spaghetti_all} further illustrate the behavior of the estimators. For each parameter combination, the 100 individual naive estimates exhibit considerable variability, especially when \(N\) is small or \(\sigma\) is large. As \(N\) increases or \(\sigma\) decreases, the spread of the naive curves narrows, indicating improved stability. The nonlinear SIMEX estimates show a similar pattern. However, their variability is generally larger than that of the naive estimates (see the variance comparison in Section~\ref{sec:discussion}). Nevertheless, the average curves demonstrate that the SIMEX correction successfully reduces bias, bringing the estimated coefficient function closer to the true function, albeit with a potential increase in variance.

\begin{figure}[htbp]
\centering
\includegraphics[width=0.85\textwidth]{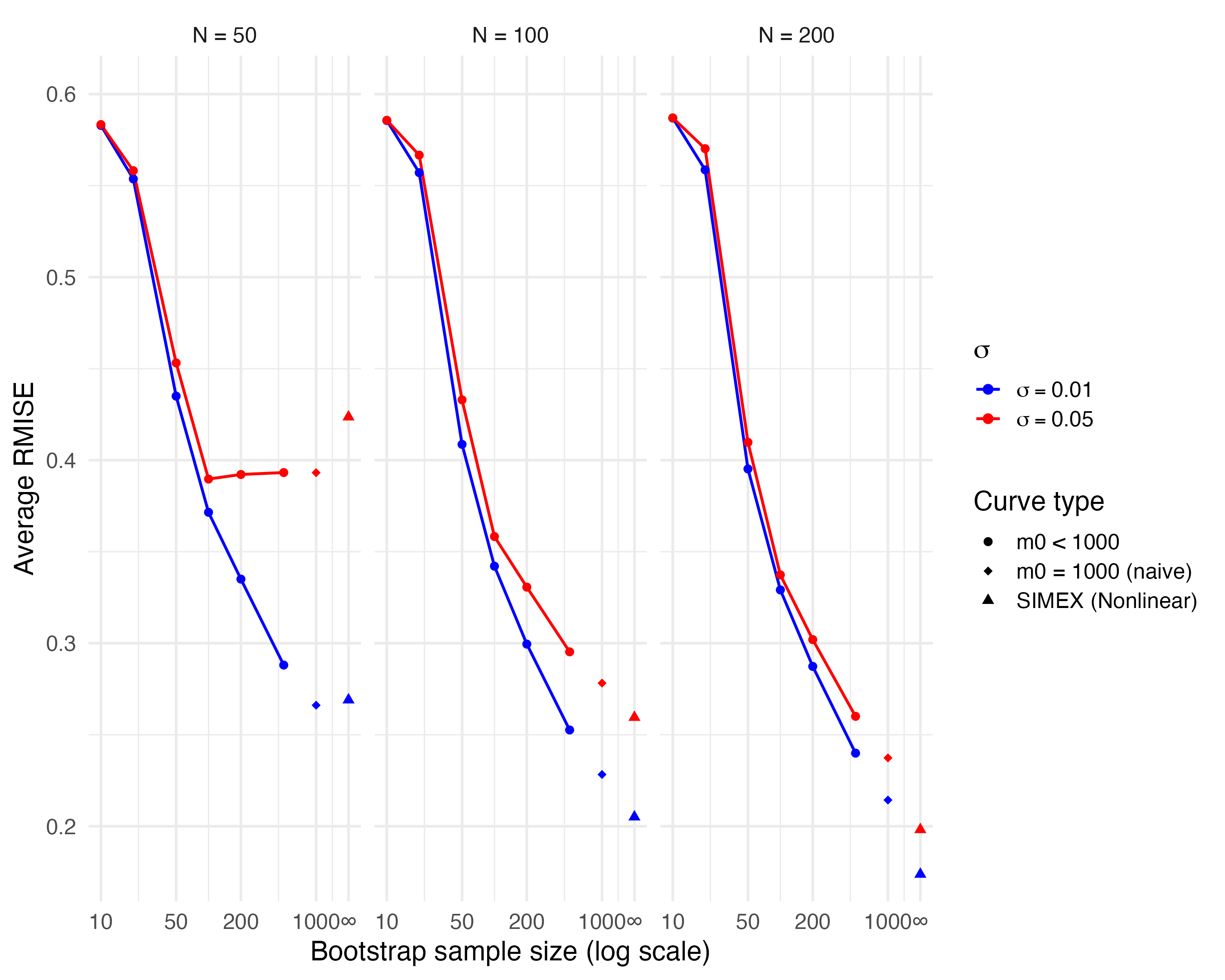}
\caption{Average RMISE of the unadjusted estimator (solid lines for \(m_0<1000\)), the naive estimator at \(m_0=1000\) (diamond), and the nonlinear SIMEX extrapolation (triangle at \(m_0=\infty\)), as functions of the bootstrap sample of size \(m_0\). Results are shown for each combination of sample size \(N\) and noise level \(\sigma\).}
\label{fig:mse_trends}
\end{figure}

\begin{figure}[htbp]
\centering
\includegraphics[width=0.85\textwidth]{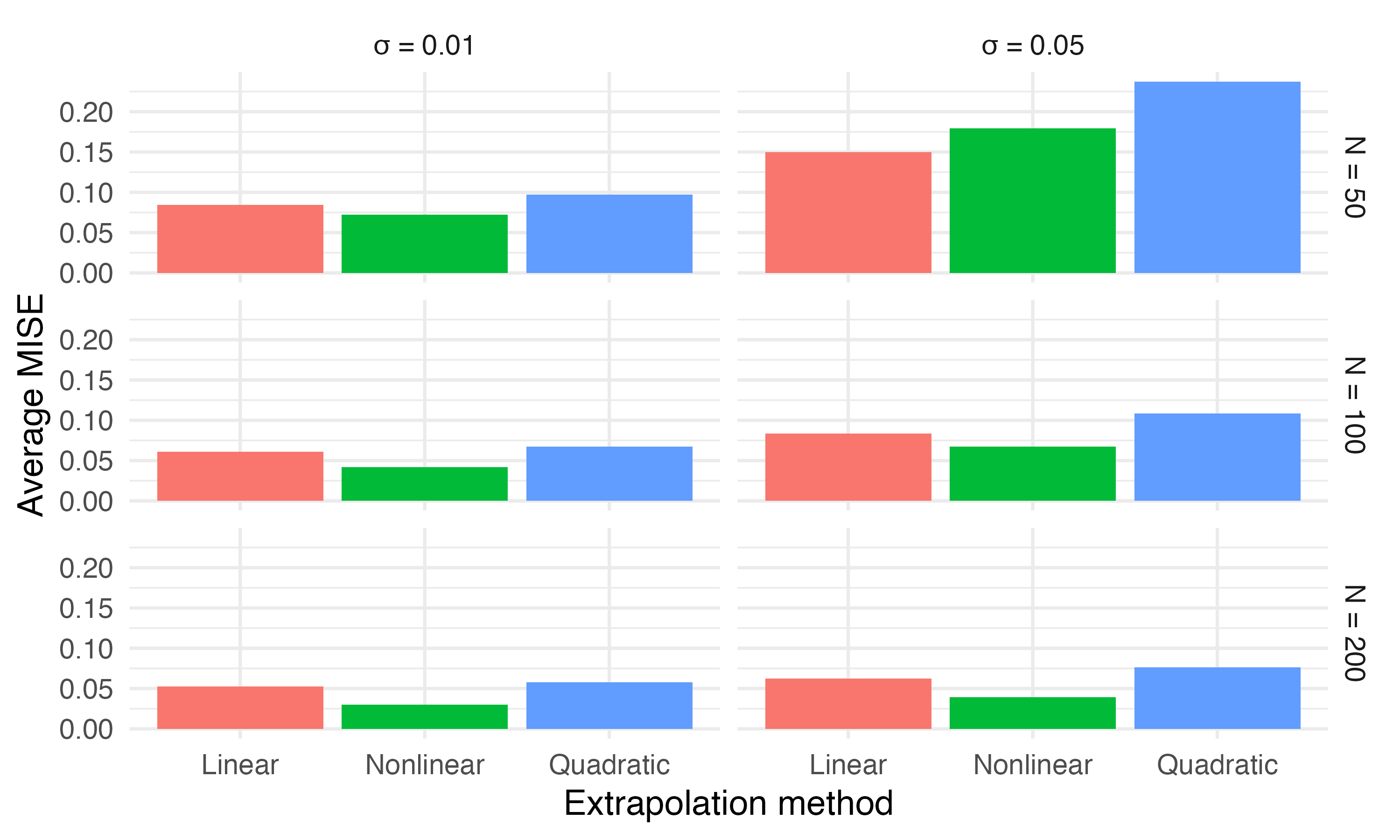}
\caption{Average MISE of the three SIMEX extrapolation methods (linear, quadratic, nonlinear) for each combination of \(N\) and \(\sigma\). The nonlinear method generally achieves the smallest MISE.}
\label{fig:mse_simex_comparison}
\end{figure}

Figure~\ref{fig:mse_trends} shows the average RMISE of the unadjusted estimator as a function of \(m_0\) for different \(N\) and \(\sigma\). For each combination, the average RMISE decreases monotonically as \(m_0\) increases, confirming that more measurements per unit lead to more accurate coefficient function estimates. The points at \(m_0=\infty\) indicate the average RMISE achieved by the nonlinear SIMEX extrapolation, which tended to be the best among the three methods (see Figure~\ref{fig:mse_simex_comparison}). In most settings, the nonlinear SIMEX estimator yields a lower average RMISE than the naive estimator based on the full set of measurements (\(m_0=1000\)), demonstrating its ability to correct for bias due to a finite number of measurements. 
An exception is seen for the smallest sample size \(N=50\), particularly for the larger variance case in which the SIMEX MISE is somewhat higher than the naive MISE (see Table~\ref{tab:simulation_metrics}) despite a clear reduction in squared bias. This is due to a substantial increase in variance, which outweighs the bias reduction. Seeing that the average RMISE of the estimates is nearly flat for all $m_0\ge 100$ suggests that, at least with this relatively high noise value, a sample size of 50 is not sufficient to gain a real benefit from the SIMEX procedure.  For larger \(N\), the variance increase is more modest, and thus the net effect is a decrease in the average RMISE.

\begin{table}[htbp]
\centering
\caption{Comparison of integrated bias², integrated variance, and MISE for naive and nonlinear SIMEX estimators.}
\label{tab:simulation_metrics}
\begin{tabular}{cccccccc}
\hline
 & & \multicolumn{2}{c}{Bias²} & \multicolumn{2}{c}{Variance} & \multicolumn{2}{c}{MISE} \\
\cline{3-4} \cline{5-6} \cline{7-8}
\(N\) & \(\sigma\) & naive & SIMEX & naive & SIMEX & naive & SIMEX \\
\hline
50  & 0.01 & 0.0496 & 0.0197 & 0.0214 & 0.0532 & 0.0710 & 0.0729 \\
50  & 0.05 & 0.1300 & 0.1120 & 0.0250 & 0.0677 & 0.1550 & 0.1800 \\
100 & 0.01 & 0.0424 & 0.0178 & 0.0098 & 0.0245 & 0.0522 & 0.0423 \\
100 & 0.05 & 0.0659 & 0.0368 & 0.0116 & 0.0308 & 0.0775 & 0.0676 \\
200 & 0.01 & 0.0414 & 0.0183 & 0.0046 & 0.0120 & 0.0460 & 0.0303 \\
200 & 0.05 & 0.0515 & 0.0259 & 0.0049 & 0.0135 & 0.0564 & 0.0394 \\
\hline
\end{tabular}
\vspace{10pt}
\end{table}

We also compared the three extrapolation methods (linear, quadratic, nonlinear) by averaging the 100 curves for each method and computing the resulting MISE. Figure~\ref{fig:mse_simex_comparison} shows that the nonlinear method generally yields the lowest MISE across all parameter combinations, confirming its superior ability to capture the curvature of the bias relationship. Consequently, we adopted the nonlinear SIMEX as the default correction method throughout.

\section{Application to NHANES Data on Physical Activity and Mortality}\label{sec:application}

We illustrate the proposed SIMEX procedure using data from the National Health and Nutrition Examination Survey (NHANES) 2011–2014 cycles \citep{akinbami2022national}. The goal is to assess the association between the 24‑hour activity profile and all‑cause mortality, while accounting for finite‑sample bias in the estimated functional predictor.

\subsection{Data Description and Preprocessing}

For each of the 12,610 participants, raw wrist-accelerometry data were processed to obtain minute-by-minute activity counts (MIMS) over a 24-hour period. For participants with multiple days of accelerometry data, the data were compressed by taking the average at each minute across available days, resulting in 1,440 observations per participant. Following standard preprocessing \citep{crainiceanu2024functional}, we restricted the analysis to participants aged 50–80 years with complete covariate and survival information, resulting in a final sample of 3,676 individuals. Among these, 429 deaths were observed during follow‑up; the remaining observations were right‑censored. For each subject we recorded either the time of death or the censoring time, together with an event indicator (1 for death, 0 for censored). Covariates included age, body mass index (BMI), gender, and coronary heart disease (CHD) status.

To avoid boundary issues in density estimation, we first divided all MIMS values by the global maximum (95.58) so that the data lie in \([0,1]\). Then, for each subject we estimated the density of their scaled activity values using smoothing splines (via \texttt{fda::density.fd}) with a fixed domain \([-0.1,1]\) to prevent near‑zero density at the left boundary. The LQD transformation was applied to obtain the functional covariate, which is defined on \(q\in[0,1]\). This transformation maps the density into the Hilbert space \(L^2[0,1]\).

\subsection{Results}

We first estimated the linear functional Cox model using the full data (all 1,440 observations per subject). Specifically, the hazard function for individual \(i\) was assumed to be
The hazard function for individual \(i\) is specified as
\[
\lambda_i(t) = \lambda_0(t)\exp\bigl\{\beta_1 X_{1i} + \beta_2 X_{2i} + \beta_3 X_{3i} + \beta_4 X_{4i} + \int_0^1 g_i(q)\beta(q)dq\bigr\},
\]
where \(\lambda_0(t)\) is the baseline hazard, \(X_{1i}\) is age (in years), \(X_{2i}\) is body mass index (BMI, in \(\mathrm{kg}/\mathrm{m}^2\)), \(X_{3i}\) is gender (1 for male, 0 for female), and \(X_{4i}\) is an indicator of coronary heart disease (CHD, 1 for yes, 0 for no). The functional covariate \(g_i(q)\) is the LQD‑transformed estimated density of the minute‑level activity data, and \(\beta(q)\) is the corresponding coefficient function. The model was fitted using the \texttt{pfr} function from the \texttt{refund} package. The functional term was specified with a $P$‑spline basis (using the \texttt{lf} constructor), setting the number of basis functions to 20 and the evaluation grid to 512 equally spaced points on \([0,1]\). A roughness penalty was applied to the integrated squared second derivative of \(\beta(q)\) to enforce smoothness, and the smoothing parameter was selected by restricted maximum likelihood (REML). The model family was set to \texttt{cox.ph} to implement the Cox proportional hazards likelihood. This estimation yielded the naive coefficient function \(\beta(q)\).

\begin{figure}[htbp]
\centering
\includegraphics[width=0.85\textwidth]{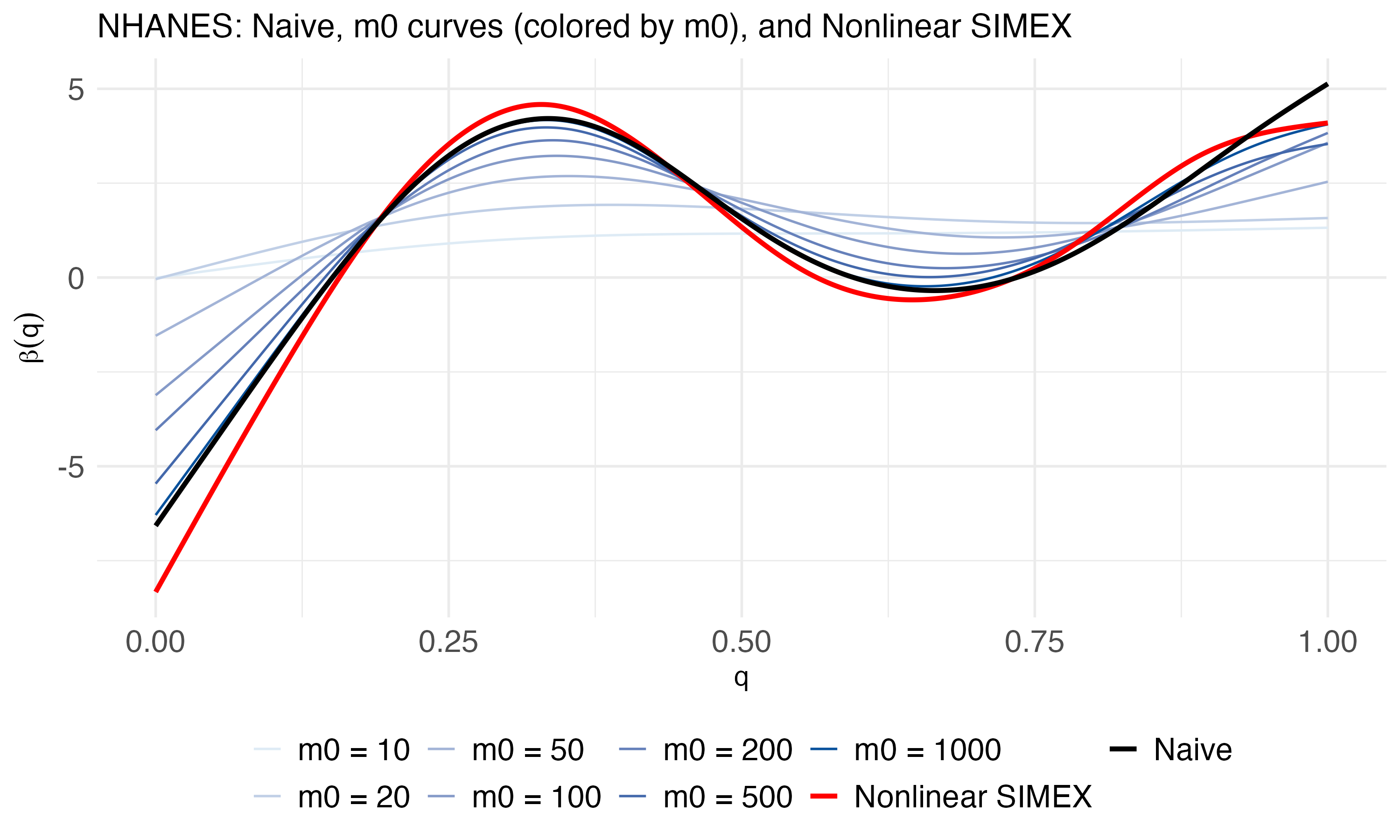}
\caption{Average coefficient functions for different bootstrap samples of size \(m_0\) (blue lines, darker for larger \(m_0\)), naive estimate (black), and nonlinear SIMEX extrapolation (red).}
\label{fig:nhanes_simex}
\end{figure}

Figure~\ref{fig:nhanes_simex} displays the results. As \(m_0\) increases, the average curves converge towards the naive estimate, suggesting that the finite‑observation bias diminishes with more observations per subject. Notably, the nonlinear SIMEX extrapolation deviates from the naive curve in a direction consistent with the simulation study, suggesting that it may be correcting some attenuation seen with the naive estimate. 
This pattern mirrors the bias reduction behavior observed in our simulations and supports the conclusion that the naive estimator displays small but non‑negligible attenuation bias even when there are 1,440 measurements per subject. 

\section{Discussion}\label{sec:discussion}

In this paper we studied the attenuation bias that arises when the functional covariate in scalar‑on‑function regression is a density estimated from a finite number of measurements per unit. Using both theoretical arguments and extensive simulations, we have shown that the bias decreases monotonically as the number of measurements per unit (\(m\)) increases, and that the proposed SIMEX procedure can effectively reduce this bias by extrapolating to the limit \(m\to\infty\). Our simulation study, conducted over a range of sample sizes \(N\) and noise levels \(\sigma\), demonstrates that the mean integrated squared error of the coefficient function declines with \(m\) and that the SIMEX‑extrapolated estimates achieve lower bias than the naive estimates based on the full set of measurements. An application to NHANES physical activity data further illustrates the method. The SIMEX correction shifts the estimate in the same direction as observed in the simulations, confirming its consistency and utility as a bias correction tool.

The SIMEX framework is agnostic to the specific density estimator used. In our implementation we employed smoothing splines (via \texttt{fda::density.fd}) with a fixed domain to avoid boundary issues, but any reasonable density estimator, such as kernel density estimation with an adaptive bandwidth, log‑spline density estimation, or even parametric models—can be plugged into the procedure. This flexibility is a key advantage. Users can choose the density estimator that best suits their data while still benefiting from the same bias‑correction strategy. The only requirement is that the estimator can be applied repeatedly to bootstrap samples of varying sizes \(m_0\).

Most existing functional SIMEX methods assume additive measurement error, typically Gaussian noise added to the observed curves \citep{cai2015methods,tekwe2022estimation, chen2024adjusting}. Our setting is fundamentally different. The error arises from estimating a density from a finite number of measurements and does not manifest itself as visible “noise” in the estimated density. Even when the estimated density appears smooth, the variability due to a finite number of measurements still causes attenuation bias in the coefficient function. A naive approach to reduce this bias would be to pre‑smooth the estimated densities before fitting the regression, but such pre‑smoothing does not eliminate the bias, and it merely trades off bias and variance. In contrast, SIMEX directly targets the bias induced by the finite‑measurement estimation of the density and has been shown to be effective even when the functional predictor is already smooth. This distinguishes our method from earlier work that compared SIMEX with simple pre‑smoothing. In our density‑regression context, pre‑smoothing is not a natural competitor because the density estimator already incorporates smoothing.

We considered three extrapolation functions: linear, quadratic, and nonlinear. In our simulations, the nonlinear method generally performed best, as it can capture the curvature of the relationship between the average coefficient function \(\bar{\hat{\beta}}_{m_0}(\cdot)\) and \(\lambda=1/\sqrt{m_0}\). However, no single method dominates in all settings. Data‑adaptive selection of the extrapolation method is possible, for example by using cross‑validation. Regarding the choice of the grid of \(m_0\) values, a practical guideline is to include values that span a wide range, from small to nearly the full number of measurements, and to examine the plot of the average coefficient function \(\bar{\hat{\beta}}_{m_0}(\cdot)\) against \(m_0\) (or \(\lambda\)). If the curves show a clear trend and begin to plateau as \(m_0\) approaches the full number of measurements, the extrapolation is likely reliable. If the trend is erratic, more \(m_0\) values or a larger maximum \(m_0\) may be needed.

\begin{figure}[htbp]
\centering
\includegraphics[width=0.85\textwidth]{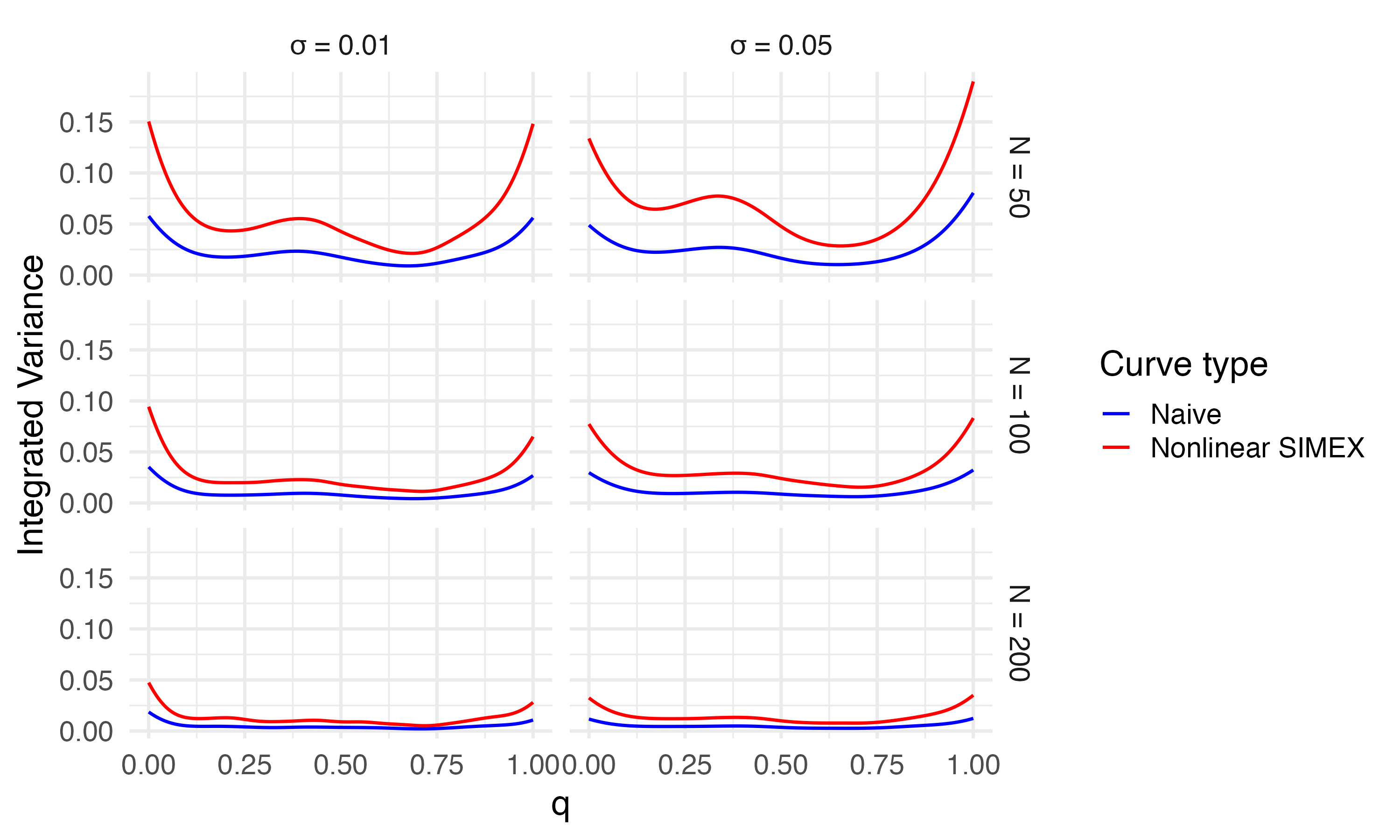}
\caption{Integrated variance of the naive (blue) and SIMEX (red) coefficient function estimates across 100 replications, for each combination of sample size \(N\) and noise level \(\sigma\).}
\label{fig:variance_curves}
\end{figure}

An additional finding from the simulation study concerns the variability of the SIMEX estimator. As shown in Figure~\ref{fig:variance_curves} and Table~\ref{tab:simulation_metrics}, the variance of the SIMEX coefficient function estimates is generally larger than that of the naive estimator based on the full number of measurements (\(m_0=1000\)). The increased variability is a natural consequence of the SIMEX procedure: the bootstrap resampling and the extrapolation steps introduce additional sampling uncertainty. This bias–variance trade‑off is well known in measurement error correction methods \citep{cook1994simulation, carroll2006measurement}. While SIMEX effectively reduces the attenuation bias (as demonstrated in Section~\ref{sec:simulation_results}), it does so at the expense of a modest increase in variance. Nevertheless, the overall improvement in MISE generally remains positive, because the reduction in squared bias outweighs the increase in variance (see Figure~\ref{fig:mse_trends}). Users of the method should be aware of this trade‑off and consider it when interpreting the results, especially in applications for which stability of the coefficient function estimates is of primary concern.

\section{Disclosure statement}\label{disclosure-statement}

The authors report there are no competing interests to declare.

\section{Data Availability Statement}\label{data-availability-statement}

The simulation data and all analysis code used to generate the results in this paper are publicly available at \url{https://github.com/XieFenglin1658/Bias-Correction-for-Scalar-on-Density-Regression-Models}. This repository contains the complete R code for generating the simulated data, applying the SIMEX procedure, and producing all figures and tables reported in the manuscript. The simulated datasets themselves are not stored due to their large size. However, they can be fully regenerated from the provided code with the specified random seeds.

The National Health and Nutrition Examination Survey (NHANES) data analyzed in Section~\ref{sec:application} are publicly available and can be accessed through the official CDC website (\url{https://wwwn.cdc.gov/nchs/nhanes/}) after free registration.

\bibliography{bibliography.bib}

\end{document}